# A STUDY OF HADRONIC BACKGROUNDS TO ISOLATED HARD PHOTON PRODUCTION WITH L3 [1]

David Kirkby
California Institute of Technology
Pasadena, CA 91125, U. S. A.
**The L3 Collaboration**

**Abstract**

I describe two methods for studying hadronic backgrounds to prompt photon production with L3, and compare the observed background rates with Monte Carlo predictions. I find that the Monte Carlo models JETSET and HERWIG underestimate the production of isolated neutral hadrons in hadronic Z decays at LEP. By extrapolating results obtained with L3, I estimate that the rate of prompt-photon + jet background to a H$\rightarrow \gamma\gamma$ search at the LHC will be larger than Monte Carlo predictions by a factor of 1.5–2.5.

[1] Work supported in part by a grant from the United States Department of Energy.

# Introduction

Energetic and isolated ("prompt") photons produced in hadronic final states originate primarily from hard scattering processes. At LEP, cuts on energy and isolation are used to study final-state radiation in hadronic Z decays[1]. At the LHC, similar cuts will be applied to select pairs of photons from the decay H$\rightarrow \gamma\gamma$[2].

Energetic and isolated neutral hadrons present a potentially large background to any study of prompt photons, since they are typically recorded in the detector as two or more overlapping photons and thus can be difficult to distinguish from single photons. The description of this background in a hadronization model is sensitive to the tails of the fragmentation function. Since these tails are poorly constrained, both theoretically and experimentally, these models are not expected to be reliable.

In this paper I use data collected with the L3 detector at LEP[3] to investigate the production of hadronic backgrounds to isolated hard photon production in hadronic Z decays. I compare the measured production rates with the predictions of the JETSET[4] and HERWIG[5] Monte Carlo models. Since present design studies for LHC detectors rely heavily on Monte Carlo predictions, I evaluate the reliability of these models for estimating the expected hadronic backgrounds to an LHC search for the decay H$\rightarrow \gamma\gamma$, based on measurements at LEP.

# Resonance Reconstruction

The first method I have used to study isolated neutral hadronic backgrounds is direct reconstruction of $\pi^0$ and $\eta$ resonances via their decay into two resolvable photons. I select photon candidates in hadronic Z decays by requiring

$$E_\gamma > 500\ \mathrm{MeV} \quad , \quad 10^\circ < \theta_\gamma < 170^\circ \quad , \quad |\Delta\phi_{\mathrm{trk}}| > 17\ \mathrm{mrad}\ ,$$

where $\Delta\phi_{\mathrm{trk}}$ is the difference in azimuthal angle between the photon candidate and the nearest charged track. I then reconstruct candidate pairs and require that the reconstructed object satisfies

$$E_{\gamma\gamma} > 3\ \mathrm{GeV} \quad , \quad 45^\circ < \theta_{\gamma\gamma} < 135^\circ\ ,$$

and that it is also isolated from other particles $\{p_i\}$ in the event by an angle $\alpha = 10^\circ$–$25^\circ$

$$\cos^{-1}(p_{\gamma\gamma} \cdot p_i) > \alpha\ .$$

Figure 1 shows the invariant mass distributions of reconstructed pairs in L3 data collected during 1991–94, with isolation requirements of $\alpha = 10^\circ$ (a) and $\alpha = 25^\circ$ (b). Narrow peaks due to $\pi^0$ and $\eta$ resonances are clearly identifiable in both cases. The JETSET Monte Carlo reproduces the observed rate of isolated $\pi^0$ and $\eta$ with $\alpha = 10^\circ$, but significantly underestimates the observed background with $\alpha = 25^\circ$.

# Shower Shape Analysis

The second method I have used to study isolated neutral hadronic backgrounds exploits the different patterns of energy deposit (*shower shapes*) in the L3 electromagnetic calorimeter due to a single photon or to multiple overlapping photons. To quantify this effect I use a neural network discriminator that classifies calorimeter clusters based on the energies measured in a $5\times5$ matrix of BGO crystals. Figure 2 shows the discriminator output distributions for clusters selected in hadronic Z decays with

$$E > 3\ \mathrm{GeV} \quad , \quad 45^\circ < \theta < 135^\circ\ ,$$

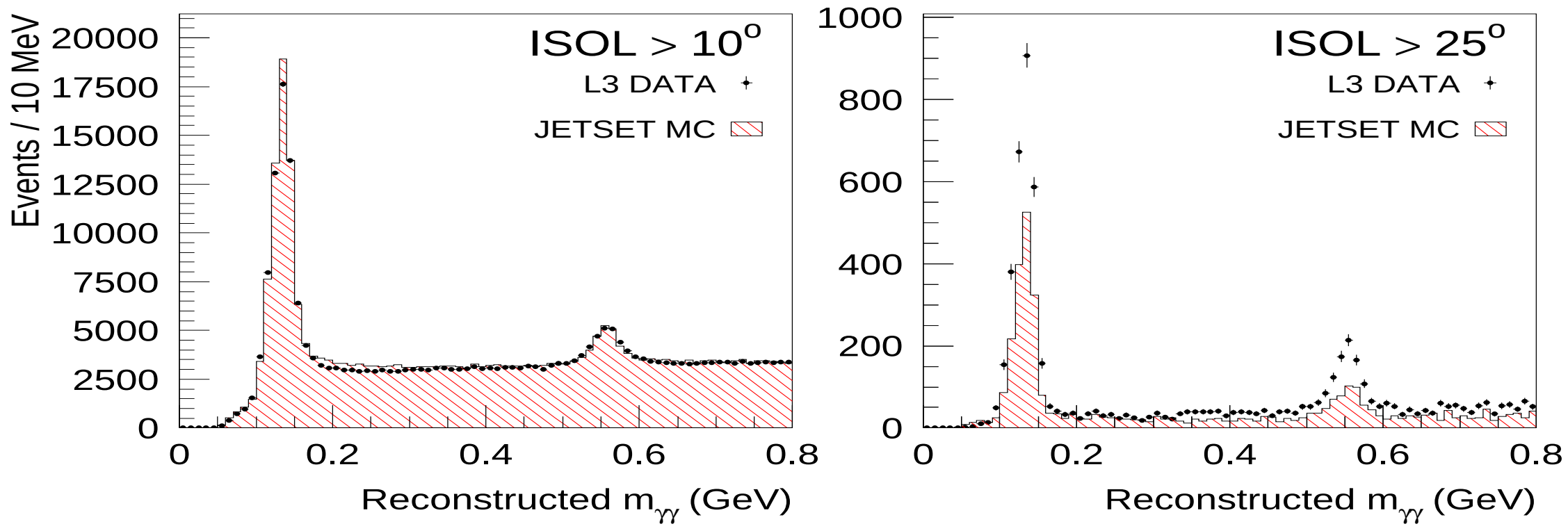

Figure 1: Invariant mass distributions for photon pairs reconstructed in hadronic Z decays, for which the reconstructed pair is isolated from other particles in the event by at least 10° (a) or 25° (b).

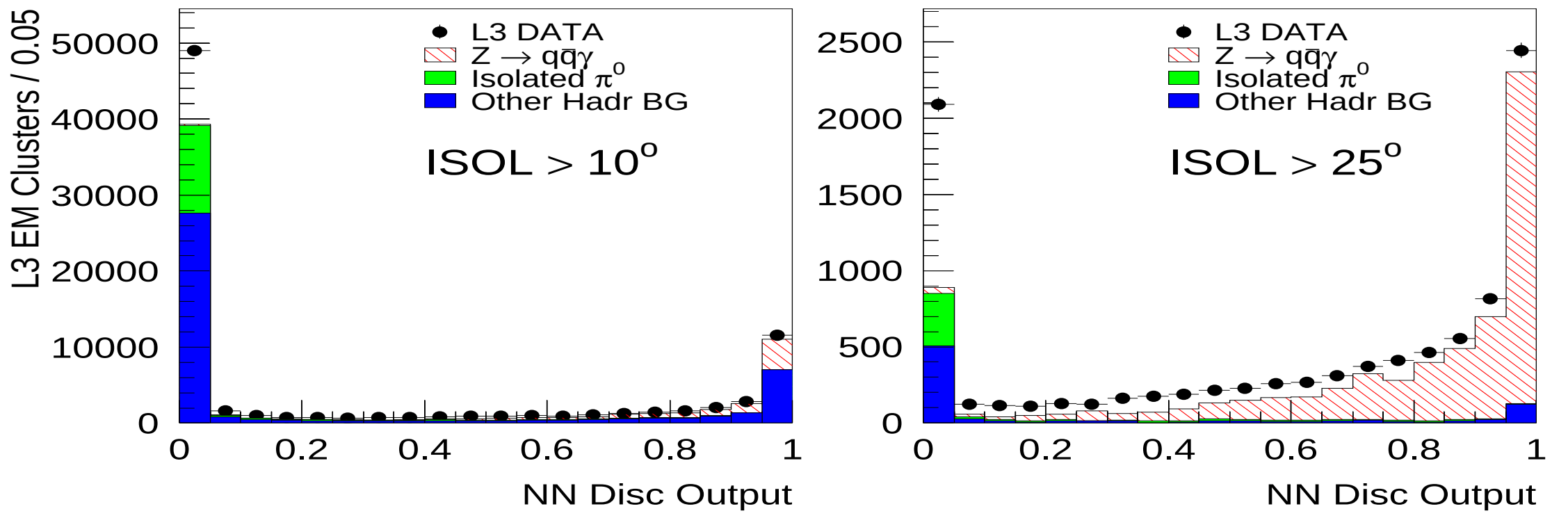

Figure 2: Neural network discriminator output distributions for electromagnetic calorimeter clusters selected in hadronic Z decays, and isolated by at least 10° (a) or 25° (b).

and isolated from other particles by $\alpha = 10°$ (a) and $\alpha = 25°$ (b). These plots show clearly that an isolation cut improves the purity of the prompt photon signal. The Monte Carlo prediction of isolated hadronic background, which is concentrated in the lowest bin, is slightly low with $\alpha = 10°$ and significantly too low with $\alpha = 25°$.

# Comparison of Methods

The two methods described above are complementary and select independent samples of hadronic background to isolated hard photon production, based on decay kinematics. At low energies, neutral hadrons usually decay into well separated photons that are selected by the first method; at intermediate energies, decay photons typically overlap in the detector and are selected by the second method; at the highest energies decays are almost collinear and thus indistinguishable from single photons.

Figure 3 gives a compilation of results obtained with the two methods described above, for an isolation requirement of $\alpha = 15°$. The two methods are found to be consistent. The disagreement between the observed and predicted background rates has the following qualitative features:

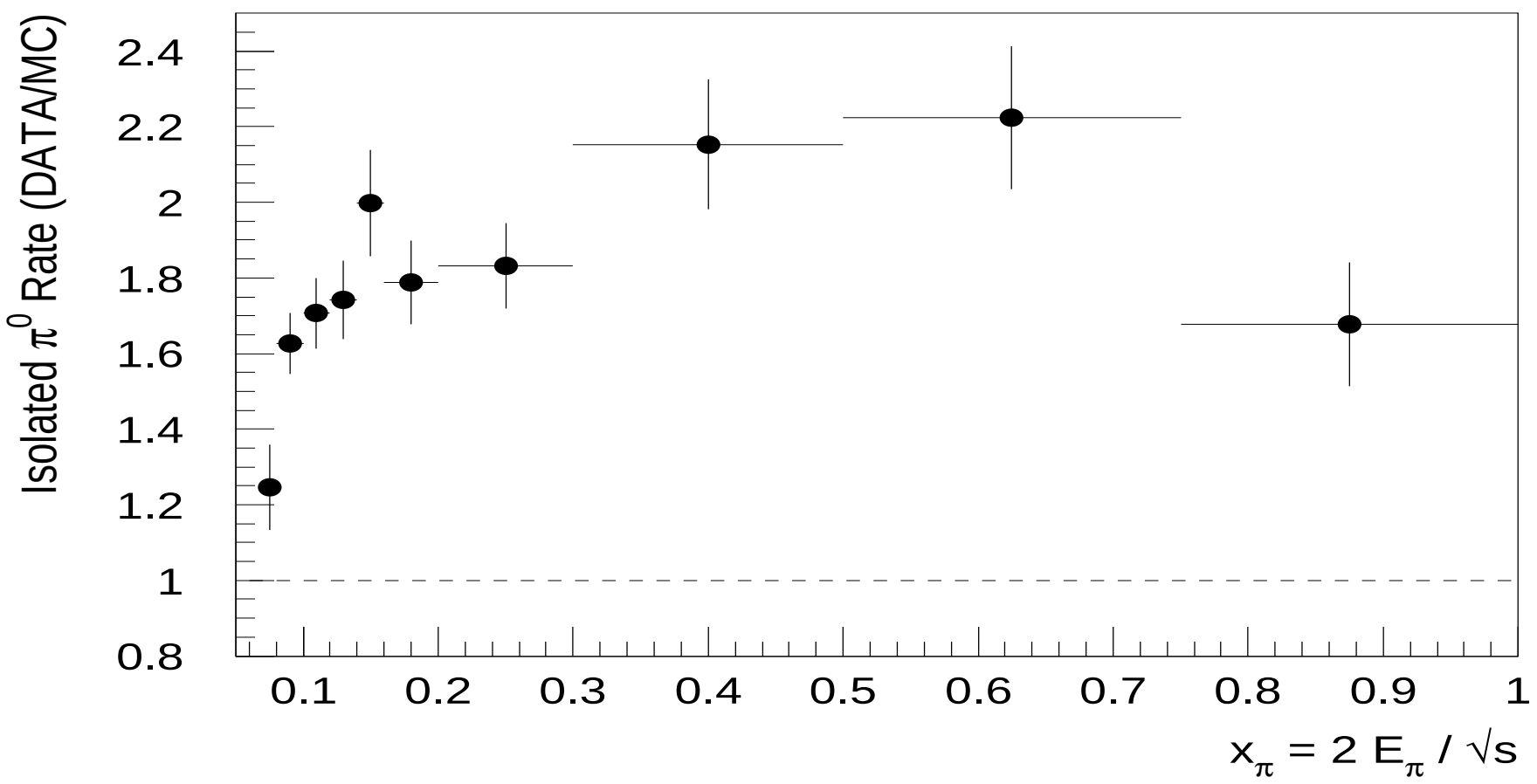

Figure 3: Compilation of measurements of the rate of isolated $\pi^0$'s in hadronic Z decays, expressed in terms of the ratio (DATA/MC) as a function of the scaled $\pi^0$ energy, $x_\pi$. Results were obtained with the JETSET Monte Carlo.

- the observed background rate is larger than the predicted rate, and this discrepancy increases with a tighter isolation cut;
- the discrepancy is largest for intermediate energies $(0.3 < x < 0.7)$;
- and the Monte Carlo model HERWIG is in slightly better agreement with data than JETSET (by 10–20%).

# Implications for Higgs Searches at the LHC

The dominant background to a search for photon pairs from Higgs decay at the LHC is expected to be from events containing a single genuine prompt photon, together with a hard isolated $\pi^0$ that is produced from a jet and then misidentified as a prompt photon. In order to perform a crude extrapolation of the backgrounds observed with L3 to the likely backgrounds to an LHC H$\rightarrow \gamma\gamma$ search, I have used PYTHIA[4] to estimate the average scaled energy $x_\pi = E_\pi/E_{\rm jet}$ of neutral pions that combine with a prompt photon to pass CMS selection cuts[2] and which also give an invariant mass in the range 90 GeV $< m_{\gamma\pi} <$ 130 GeV, where the two photon Higgs decay mode is expected to be competitive; I find $\langle x_\pi \rangle \simeq 0.4$. The proposed CMS isolation scheme in azimuthal angle and pseudorapidity corresponds approximately to $\alpha = 15^\circ$ at LEP. By combining these results, I estimate that the prompt photon + jets background to an LHC H$\rightarrow \gamma\gamma$ search will be larger than Monte Carlo predictions by a factor of 1.5–2.5.